\documentclass{ws-ijgmmp}
\usepackage[utf8]{inputenc}
\begin{document}

\markboth{Piyali Bhar et al.}
{Finch-Skea star model in $f(R,T)$ theory of gravity}

%
\catchline{}{}{}{}{}
%

\title{Finch-Skea star model in $f(R,T)$ theory of gravity}

\author{Piyali Bhar}
\address{Department of
Mathematics,Government General Degree College, Singur, Hooghly, West Bengal 712 409,
India \\
\email{piyalibhar90@gmail.com, piyalibhar@associates.iucaa.in
 }}

\author{Pramit Rej}
 \address{Department of
Mathematics, Sarat Centenary College, Dhaniakhali, Hooghly, West Bengal \\ 712 302, India \\
\email{pramitrej@gmail.com
 }}

\author{Aisha Siddiqa}
\address{Department of Mathematics, Virtual University of Pakistan,\\
54-Lawrence Road, Lahore, Pakistan.\\
\email{aisha.siddiqa@vu.edu.pk}}

\author{Ghulam Abbas}
\address{Department of Mathematics, Islamia University Bahawalpur,\\
Pakistan \\
\email{abbasg91@yahoo.com}}

\maketitle

\begin{history}
\received{(Day Month Year)}
\revised{(Day Month Year)}
\end{history}

\begin{abstract}
The present work discusses about the existence of compact star model in the
context of $f(R,T)$ gravity with $R$ as the Ricci scalar and $T$ as the
trace of energy-momentum tensor $T_{\mu \nu}$. The model has been
developed by considering the spherically symmetric spacetime
consisting of isotropic fluid with $f(R,T)=R+2 \beta T$ with $\beta$
be the coupling parameter. The corresponding field equations are solved
by choosing well known Finch-Skea {\em ansatz} [Finch, M.R., Skea, J.E.F.:{\it Class. Quantum Gravity} {\bf 6}, 467 (1989)]. For spacetime
continuity we elaborate the boundary conditions by considering the
exterior region as Schwarzschild metric. The unknown constants
appearing in the solution are evaluated for the compact star PSR~J 1614-2230 for different values of coupling constant. The
physical properties of the model, e.g., matter density, pressure,
stability etc. have been discussed both analytically and graphically. This analysis showed
that the geometry and matter are compatible with each other as well
as the model is in stable equilibrium in the context of $f(R,\,T)$ modified gravity.
\end{abstract}

\keywords{Compact objects, $f(R,T)$ gravity, stability}



\section{Introduction}

The motivation behind the extended theories of gravity is mainly the
issue of so called dark energy which is causing the universe to
expand at an accelerated rate and also due to the unknown behavior of
gravity at quantum level. Many alternative approaches to general
relativity (GR) have been proposed in literature inducing changes in
curvature or matter aspects to explore these issues on theoretical
grounds. Among these theories, the $f(R)$ gravity is simplest one.
Furthermore, the curvature matter coupling theories are mainly
developed to study the gravity at quantum level. $f(R,T)$
theory was developed by Harko et al. \cite{Harko:2011kv}. This modified
theory of gravity has gained much attention to explore the various
astrophysical as well as cosmological conjectures since last few years.

Compact objects are formed after the supernova explosion at the end
of a star's life. These objects including black holes, neutron stars
and white dwarfs are very dense as compared to the star forming
these objects. Neutron stars are the densest observable structures
in the universe. They can be used to study all the four fundamental
interactions among particles at extreme dense level. Moreover, the
matter in massive neutron star have the possibility to decompose
into quarks and ultimately the star becomes quark star. This
conversion of neutron stars into quark stars has been discussed by several
researchers. Benvenuto and Lugones \cite{Benvenuto:1999uk} discussed the
transformation of nuclear matter into quark matter in evolving
proto-neutron star. Annala et al. \cite{Annala:2019puf} reported the evidence
for the presence of quark matter in the cores of massive neutron
stars. Odintsov and Oikonomou \cite{Odintsov:2021nqa} investigated the effects
of the Higgs model on static neutron stars, particularly they derived the Einstein frame Tolman-Oppenheimer-Volkoff equations by numerically integrating them for both the interior and the
exterior of the neutron star. Odintsov and Oikonomou \cite{Odintsov:2021qbq} also studied the implications of a subclass of E-models cosmological attractors,
namely of a-attractors, on hydrodynamically stable slowly rotating neutron stars and presented
the Jordan frame theory of the a-attractors and by using a conformal transformation
they derived the Einstein frame theory.

The significant characteristics of these compact objects are the
source of motivation for researchers to discuss their properties as
well as internal formation in various frameworks. They adopt
different ways and ingredients to study these compact objects
theoretically. The observed compact star candidates, equations of
state as well as constraints on metric potentials are the main tools
of current research in general relativity and modified theories as
well. Harko et al. \cite{Harko:2013wka} examined the structural properties
of some neutron, quark and exotic stars in Eddington-inspired
Born-Infeld (EiBI) gravity. They concluded that in EiBI gravity star
are more massive than GR for various equations of state. Nojiri et al. \cite{Nojiri:2017ncd} presented the formalism of standard modified gravity theory representatives, like
$f(R)$, $f(\mathcal{G})$ and $f(\mathbb{T})$ and other several alternative theoretical proposals which
appeared in the literature during the last decade and they explained how these theories can be considered as viable descriptions for our Universe.
Astashenok et al. \cite{Astashenok:2021peo} investigated the causal limit of maximum mass for stars in the framework of $f(R)$ gravity by choosing a causal equation of state, with variable speed of sound, and with the transition density
and pressure corresponding to the SLy equation of state. Astashenok et al. \cite{Astashenok:2020qds} studied that a neutron star with a mass in the range $(2.50- 2.67)~M_{\odot}$ can be consistently explained with the mass-radius relation in the context of Extended Theories of Gravity.
Furthermore, the authors adopted the equations of state consistent with LIGO observational constraints and concluded that
the masses of rotating neutron stars can exceed $2.6~M_{\odot}$ for some equations of
state compatible with LIGO data. Lau et al.
\cite{Lau:2017qtz} studied the tidal deformation of compact configurations
with crystalline quark matter. Deb et al. \cite{Deb:2017rhc} studied the
anisotropic quark stars in $f(R,T)$ theory. They found that with an
increase in coupling parameter mass and radius of compact star
increase and correspondingly its compactness decreases. Lopos and
Panotopoulos \cite{Lopes:2018oao} investigated the characteristics of
compact stars consisting of a mixture of dark matter and strange
quark matter. They consider the MIT bag model and polytropic EoS for
quark matter and dark matter, respectively finding that the effects
of dark matter and modified gravity is same on quark stars. Abbas et al. \cite{Abbas:2015wea} examined the existence of stable strange
stars in $f(\mathbb{T})$ theory, where $\mathbb{T}$ stands for
torsion. They analyzed their physical characteristics via MIT bag
EoS. The equilibrium state of these compact objects is investigated
in \cite{Moraes:2015uxq} for above mentioned equations of state models in
$f(R,T)$ theory. Sharif and Siddiqa \cite{Sharif:2018hoy} investigated the
outcomes of additional terms due to curvature matter coupling model
$R+\alpha R^{2} + \lambda T$ on the features of compact stars
obeying polytropic EoS and MIT bag EoS. Sharif and Waseem \cite{Sharif:2018khl}
studied the quark stars structure in $f(R,T)$ context constraining
the matter via MIT bag model and geometry by Krori-Barua solution.
 Biswas et al. \cite{Biswas:2020gzd} studied the spherically
symmetric quark star models under anisotropic matter distribution in
$f(R,T)$ framework. They considered the Tolman-Kuchowicz type metric
potentials and discussed the different properties of quark stars. Recently Bhar \cite{Bhar:2020abv} proposed a model of compact star in $f(R,T)$ gravity by employing the Krori-Barua ansatz. Zubair et al. \cite{Zubair:2021mba} proposed a model in $f(R,\,T)$ gravity using karmarkar condition. Rej and Bhar \cite{Rej:2021ngp} proposed strange stars model in the framework of $f(R,T)$ theory of gravitation in presence of charge by employing the Krori-Barua {\em ansatz}. The evolution of spherically symmetric charged anisotropic viscous fluids is discussed in framework of $f(R,T)$ gravity was studied by Noureen et al \cite{Noureen:2021xlf}. Zubair and Azmat \cite{Zubair:2020poe} have studied a cylindrically symmetric self-gravitating dynamical object via complexity factor which is obtained through orthogonal splitting of Reimann tensor in $f(R,T)$ theory of gravity. Bhatti et al. \cite{Bhatti:2020rzr} studied the stability of axially symmetric compact system with anisotropic environment in the background of $f(R,T)$ gravity. Rej et al. \cite{Rej:2021qpi} obtained a model of compact star within the framework of $f(R,\,T)$ modified gravity theory using the metric potentials proposed by Tolman-Kuchowicz corresponding to the exterior Reissner–Nordstr\"{o}m line element.\par

In present paper, we use Finch-Skea \cite{Finch_1989} metric potential to study the compact stellar
structures in extended theory of gravity. Finch and skea \cite{Finch_1989} proposed a new realistic model for
spherical symmetric stellar configurations. This Finch-Skea {\em ansatz} has been
considered in literature by several authors. Bhar \cite{Bhar:2015qza} worked out the physical
properties of strange stars admitting Chaplygin gas EoS and Finch-Skea metric potentials. In \cite{pandya2015modified}, the authors considered a
generalized Finch-Skea metric potential along with a choice of
radial pressure for a spherically symmetric anisotropic stellar
configuration. They showed that a tuneable parameter makes possible
for many observed pulsars to be accommodated in this model. Maharaj
et al. \cite{maharaj2017family} solved the Einstein-Maxwell field equations
and found a family of exact solutions in terms of Bessel functions. The purpose of this study
is to present the physical analysis of a spherically symmetric
compact star model admitting the limitations of Finch Skea geometry
in the coupling background. We consider the simplest non-minimal
coupling between matter and geometry described by the model
$f(R,T)=R+2\beta T$ and also noted the effects of coupling parameter
$\beta$ on the physical features of compact stars. \par

We have organized our paper as follows : in the next section we have described the basic field equations in $f(R,\,T)$ gravity.
Section \ref{sec3} gives the solution of field equations using Finch-Skea model. In section \ref{sec4} we have matched our interior spacetime to the exterior Schwarzschild line element to the boundary of the star.
Section \ref{pa} describes the physical properties of the present model both analytically and graphically. The next section provides the stability of the present model and some concluding remarks are given in section \ref{conclusion}.
\section{Basic Equations}\label{sec2}

The action of $f(R,T )$ gravity as proposed by Harko {\em et al.}
\cite{Harko:2011kv} is,
\begin{eqnarray}\label{action}
S&=&\frac{1}{16 \pi}\int  f(R,T)\sqrt{-g} d^4 x + \int \mathcal{L}_m\sqrt{-g} d^4 x,
\end{eqnarray}
where $f ( R,T )$ represents the general function of Ricci scalar
$R$ and trace $T$ of the energy-momentum tensor $T_{\mu \nu}$,
$\mathcal{L}_m$ being the Lagrangian matter density and $g =
det(g_{\mu \nu}$). The energy momentum tensor of matter is defined
by \cite{landau2013classical},
\begin{eqnarray}\label{tmu1}
T_{\mu \nu}&=&-\frac{2}{\sqrt{-g}}\frac{\delta \sqrt{-g}\mathcal{L}_m}{\delta \sqrt{g_{\mu \nu}}},
\end{eqnarray}
and its trace is given by $T=g^{\mu \nu}T_{\mu \nu}$. If the Lagrangian matter density $\mathcal{L}_m$ depends only on $g_{\mu \nu}$, not on its derivatives, eqn.(\ref{tmu1}) becomes,
\begin{eqnarray}
T_{\mu \nu}&=& g_{\mu \nu}\mathcal{L}_m-2\frac{\partial \mathcal{L}_m}{\partial g_{\mu \nu}}.
\end{eqnarray}

The general field equation for action given in (\ref{action}) is
obtained as,
\begin{eqnarray}\label{frt}
f_R(R,T)R_{\mu \nu}-\frac{1}{2}f(R,T)g_{\mu \nu}+(g_{\mu \nu }
\Box-\nabla_{\mu}\nabla_{\nu})f_R(R,T)&=&8\pi T_{\mu \nu}-
f_T(R,T)T_{\mu \nu}\nonumber\\&&-f_T(R,T)\Theta_{\mu \nu},
\end{eqnarray}
with, $f_R(R,T)=\frac{\partial f(R,T)}{\partial
R},~f_T(R,T)=\frac{\partial f(R,T)}{\partial T}$. $\nabla_{\nu}$
represents the covariant derivative associated with the Levi-Civita
connection of $g_{\mu \nu}$, $\Theta_{\mu \nu}=g^{\alpha
\beta}\frac{\delta T_{\alpha \beta}}{\delta g^{\mu \nu}}$ and $\Box
\equiv \frac{1}{\sqrt{-g}}\partial_{\mu}(\sqrt{-g}g^{\mu
\nu}\partial_{\nu})$ represents the D'Alembert operator.\\
The divergence of $T_{\mu \nu}$ is given (For details see ref
\cite{Harko:2011kv} and\cite{Koivisto:2005yk}) as,
\begin{eqnarray}\label{conservation}
\nabla^{\mu}T_{\mu \nu}&=&\frac{f_T(R,T)}{8\pi-f_T(R,T)}
\left[(T_{\mu \nu}+\Theta_{\mu \nu})\nabla^{\mu}\ln f_T(R,T)+\nabla^{\mu}\Theta_{\mu \nu}\right].
\end{eqnarray}

From eqn.(\ref{conservation}), we can check that $\nabla^{\mu}T_{\mu
\nu}\neq 0$ if $f_T(R,T)\neq 0.$ In particular the equations of $f(R)$ theory are obtained for
$f(R,T)=f(R)$. Furthermore, for our present model, the following
separable form of curvature matter coupling function is considered
\begin{eqnarray}\label{e}
f(R,T)&=& R+2 \beta T,
\end{eqnarray}
where $\beta$ is a small positive constant.\\
Let us consider the static spherical symmetric line
element :
\begin{equation}\label{line}
ds^{2}=-e^{\nu(r)}dt^{2}+e^{\lambda(r)}dr^{2}+r^{2}\Big[\sin^{2}\theta d\phi^{2}+d\theta^{^2}\Big],
\end{equation}
to describe a spherically symmetry $4D$ spacetime in Schwarzschild coordinates $x^{\mu}=(t,\,r,\,\theta,\,\phi)$.
We also assume that the matter in interior structure is perfect fluid with energy momentum tensor
\begin{eqnarray}
T_{\mu \nu}&=&(\rho+p)u_{\mu}u_{\nu}-p g_{\mu \nu},
\end{eqnarray}
where $\rho$ is the matter density and $p$ is the isotropic pressure in
modified gravity, $u^{\mu}$ is the fluid four velocity satisfies the
equations $u^{\mu}u_{\mu}=1$ and $u^{\mu}\nabla_{\nu}u_{\mu}=0$. In our present work
the matter Lagrangian can be taken as $\mathcal{L}_m=-p$ following
Harko et al.\cite{Harko:2011kv}. The expression of $\Theta_{\mu
\nu}$ is given by, $-2T_{\mu \nu}-pg_{\mu\nu}$.
Moreover, replacing (\ref{e}) into (\ref{frt}) one can obtain,
\begin{eqnarray}
G_{\mu \nu}&=&8\pi T_{\mu \nu}^{\text{eff}},
\end{eqnarray}
where $G_{\mu \nu}$ is the Einstein tensor and
\begin{eqnarray}
T_{\mu \nu}^{\text{eff}}&=& T_{\mu \nu}+\frac{\beta}{8\pi}
T g_{\mu \nu}+\frac{\beta}{4\pi}(T_{\mu \nu}+p g_{\mu \nu}).
\end{eqnarray}
For metric given in (\ref{line}), the above equation produces the
following field equations,
\begin{eqnarray}
8\pi\rho^{\text{eff}}&=&\frac{\lambda'}{r}e^{-\lambda}+\frac{1}{r^{2}}(1-e^{-\lambda}),\label{f1}\\
8 \pi p^{\text{eff}}&=& \frac{1}{r^{2}}(e^{-\lambda}-1)+\frac{\nu'}{r}e^{-\lambda},\label{f2} \\
8 \pi p^{\text{eff}}&=&\frac{1}{4}e^{-\lambda}\left[2\nu''+\nu'^2-\lambda'\nu'+\frac{2}{r}(\nu'-\lambda')\right], \label{f3}
\end{eqnarray}
where $\rho^{\text{eff}}$ and $p^{\text{eff}}$ represent the density
and pressure, respectively, in Einstein Gravity and
\begin{eqnarray}
\rho^{\text{eff}}&=& \rho+\frac{\beta}{8\pi}(3 \rho-p),\label{r1}\\
p^{\text{eff}}&=& p-\frac{\beta}{8\pi}(\rho-3p).\label{r2}
\end{eqnarray}
Here prime corresponds to a derivative with respect to radial
coordinate. In the above two equations, $\rho$ and $p$ respectively denote the matter density and pressure in modified gravity.\\ Using Eqs. (\ref{f1})-(\ref{f3}), with the help of (\ref{r1}) and (\ref{r2}), we get,
\begin{eqnarray}\label{con}
\frac{\nu'}{2}(\rho+p)+\frac{dp}{dr}&=&\frac{\beta}{8\pi+2\beta}(p'-\rho').
\end{eqnarray}
The above equation is the TOV equation in modified gravity.\\
In coming section, we proceed to solve the eqns.
(\ref{f1})-(\ref{f3}) to obtain the compact star model.

\section{Finch-Skea metric potential and proposed model}\label{sec3}

In eqns. (\ref{f1})-(\ref{f3}), there are three eqns. in four
unknowns. So in order to solve the system we have to fix any one.
There are different ways to eliminate one unknown, however, here we fix $g_{rr}$ by the following expression,
\begin{eqnarray}\label{elambda}
e^{\lambda}&=&1+ar^2,
\end{eqnarray}
where $a$ is a constant of dimension km$^{-2}$. This metric potential was proposed by Finch and Skea \cite{Finch_1989} and it is familiar as Finch-Skea {\em ansatz}. Bhar et al. \cite{Bhar:2014iya} proposed a new class of
interior solutions of a ($2 + 1$)-dimensional anisotropic star
in a Finch-Skea spacetime corresponding to the exterior
BTZ black hole. A relativistic stellar model admitting a quadratic equation of state was proposed by Sharma and Ratanpal \cite{Sharma:2013lqa} by taking Finch-Skea {\em ansatz}. Zubair et al. \cite{Zubair:2021lhe} studied the spherically symmetric compact star model with anisotropic matter distribution in the framework of $f(T)$ modified gravity by taking Finch-Skea metric. Shamir et al. \cite{FarasatShamir:2021mmi} explored the compact geometries by employing Karmarkar condition with the charged anisotropic source of matter distribution by taking Finch-Skea geometry. Bhar et al. \cite{Bhar:2017jbl} discovered a new well-behaved charged anisotropic solution of Einstein-Maxwell's field equations under embedding class-I in the background of Finch-Skea geometry.\par

Plugging (\ref{elambda}) into (\ref{f1}), we obtain,
\begin{eqnarray}
\rho^{\text{eff}}&=&\frac{a (3 + a r^2)}{8 \pi (1 + a r^2)^2}
\end{eqnarray}
From eqns. (\ref{f2})-(\ref{f3}), we get,
\begin{eqnarray}\label{c}
r^2(2\nu''+\nu'^2-\nu'\lambda')-2r(\nu'+\lambda')+4(e^{\lambda}-1)=0.
\end{eqnarray}
Solving (\ref{c}) with the help of (\ref{elambda}), we obtain,
\begin{eqnarray}
e^{\nu}&=&\big((B - A Z) \cos Z + (A + B Z) \sin Z\big)^2,
\end{eqnarray}
where $Z=\sqrt{1 + a r^2}$. The expression for $p^{\text{eff}}$ is thus obtained as,
\begin{eqnarray}
p^{\text{eff}}&=&\frac{a \Big((B + A Z) \cos Z + (A - B Z) \sin Z\Big)}{8 \pi Z^2 \Big((B - A Z) \cos Z + (A + B Z) \sin Z\Big)}.
\end{eqnarray}
Using the expression of $p^{\text{eff}}$ and $\rho^{\text{eff}}$, from eqns. (\ref{r1}) and (\ref{r2}), we obtain the expression of matter density and pressure $\rho,\,p$ in modified gravity as,

\begin{eqnarray}
\rho&=&\frac{a}{8 (\beta + 2 \pi) (\beta + 4 \pi) Z^4}\bigg[(3 \beta + 8\pi) (3 + a r^2)+\frac{\beta Z^2 \Psi_1}{\Psi_2}\bigg],\label{p1}\\
p&=&\frac{a}{8 (\beta + 2 \pi) (\beta + 4 \pi) Z^4}\bigg[\beta (3 + a r^2)+\frac{(3 \beta + 8 \pi) Z^2 \Psi_1}{\Psi_2}\bigg].\label{p2}
\end{eqnarray}
where, $\Psi_1$ and $\Psi_2$ are functions of $r$ and their expressions are given as,
\begin{eqnarray*}
\Psi_1&=&(B + A Z) \cos Z + (A - B Z) \sin Z,\\
\Psi_2&=&(B - A Z) \cos Z + (A + B Z) \sin Z.
\end{eqnarray*}


The expressions of central density and pressure in the assumed
scenario are obtained as,
\begin{eqnarray*}
\rho_c &=&\frac{a \Big(4 A (\beta + 3 \pi) - B (5 \beta + 12 \pi)\Big) \cos 1 -
 a \Big((5 A + 4 B) \beta + 12 (A + B) \pi\Big) \sin 1}{4 (\beta + 2 \pi) (\beta + 4 \pi) \Big((A - B) \cos 1 - (A + B) \sin 1\Big)},\\
 p_c&=&\frac{-a \Big(3 B \beta + 4 (A + B) \pi\Big) \cos 1 +
 a \Big(-3 A \beta - 4 A \pi + 4 B \pi\Big) \sin 1}{4 (\beta + 2 \pi) (\beta + 4 \pi) \Big((A - B) \cos 1 - (A + B) \sin 1\Big)}.
\end{eqnarray*}
clearly the model does not suffer from the central singularities since $\rho_c$ and $p_c$ are finite.

\section{Boundary Condition}\label{sec4}

In this section, the matching of interior and exterior line elements
is given. The exterior spacetime is given by Schwarzschild metric as:
\begin{eqnarray}
ds_+^{2}&=&-\left(1-\frac{2M}{r}\right)dt^{2}+\left(1-\frac{2M}{r}\right)^{-1}dr^{2}+r^{2}\left(d\theta^{2}
+\sin^{2}\theta d\phi^{2}\right).
\end{eqnarray}
Now, at the boundary $r=r_b$, the interior spacetime is given by the line element,
 \begin{eqnarray}
ds_{-}^2 & =&-\big((B - A Z) \cos Z + (A + B Z) \sin Z\big)^2 dt^2+ \left(1 + ar^2\right) dr^2\nonumber\\&&+r^2(d\theta^2+\sin^2 \theta d\phi^2),
\end{eqnarray}
A smooth matching of metric potentials require that across the
boundary $r=r_b$,
\begin{eqnarray}\label{b1}
g_{rr}^+=g_{rr}^-,~g_{tt}^+=g_{tt}^-,
\end{eqnarray}
and the second fundamental form implies
\begin{eqnarray}\label{b2}
p(r=r_b-0)=p(r=r_b+0).
\end{eqnarray}
The boundary conditions $g_{tt}^+=g_{tt}^-$ and $p(r_b)=0$, yield
\begin{eqnarray}
1-\frac{2M}{r_b}&=&(1+ar_b^2)^{-1},\label{o1}\\
1-\frac{2M}{r_b}&=& \big((B - A Z(r_b)) \cos Z(r_b) + (A + B Z(r_b)) \sin Z(r_b)\big)^2.\label{o2}
\end{eqnarray}
\begin{itemize}
  \item {\bf Determination of $a$:} ~Now the boundary condition $g_{rr}^+=g_{rr}^-$, implies,
\begin{eqnarray}
a &=& -\frac{2 M}{(2 M - r_b) r_b^2}\label{b1}
\end{eqnarray}
  \item {\bf Determination of $A$ and $B$ :}~ Solving the equations (\ref{o1}) and (\ref{o2}), we obtain the expressions for $A$ and $B$ as,
  \begin{eqnarray}
 B&=&\frac{-4 A \pi \sqrt{1 + a r_b^2} \cos \sqrt{1 + a r_b^2} -
 a A \beta r_b^2 \sqrt{1 + a r_b^2} \cos \sqrt{1 + a r_b^2}-C_1}{3 \beta \cos \sqrt{1 + a r_b^2} + 4 \pi \cos \sqrt{1 + a r_b^2} +
 2 a \beta r_b^2 \cos \sqrt{1 + a r_b^2} +C_2},
 \\
  A&=& \frac{-\sqrt{1-\frac{2M}{r_b}}+B\cos\sqrt{\frac{r_b}{r_b-2M}}+B\sqrt{\frac{r_b}{r_b-2M}}\sin\sqrt{\frac{r_b}{r_b-2M}}}{\sqrt{\frac{r_b}{r_b-2M}}\cosh\sqrt{\frac{r_b}{r_b-2M}}-\sin\sqrt{\frac{r_b}{r_b-2M}}}.
 \end{eqnarray}
The expressions of the constants $C_1$ and $C_2$ are given by,
\begin{eqnarray*}
C_1&= &
 4 a A \pi r_b^2 \sqrt{1 + a r_b^2} \cos \sqrt{1 + a r_b^2} -
 3 A \beta \sin \sqrt{1 + a r_b^2} - 4 A \pi \sin \sqrt{1 + a r_b^2} \\&&-
 2 a A \beta r_b^2 \sin \sqrt{1 + a r_b^2} -
 4 a A \pi r_b^2 \sin \sqrt{1 + a r_b^2},\\
 C_2&=&4 a \pi r_b^2 \cos \sqrt{1 + a r_b^2} -
 4 \pi \sqrt{1 + a r_b^2} \sin \sqrt{1 + a r_b^2} \\&&-
 a \beta r_b^2 \sqrt{1 + a r_b^2} \sin \sqrt{1 + a r_b^2} -
 4 a \pi r_b^2 \sqrt{1 + a r_b^2} \sin \sqrt{1 + a r_b^2}.
 \end{eqnarray*}
 \end{itemize}
 The numerical values of $a,\,A$ and $B$ for different values of $\beta$ are obtained in table~\ref{table1}.

 \begin{table*}[t]
\centering
\caption{The values of the constants $a,\,A$ and $B$ for the compact star PSR J1614-2230.}
\label{table1}
\begin{tabular}{@{}cccccccccccccccc@{}}
\toprule
Objects  & Estimated &Estimated & $\beta$ && $a$&& $A$  &&$B$ \\
&Mass ($M_{\odot}$)& Radius &&& (km$^{-2}$) &\\
\hline
PSR J1614-2230 \cite{Gangopadhyay:2013gha}&$1.97$&$9.69$& 0.0&& $0.01596$ && $0.318282 $ && $0.197009$\\
&&& $0.3$ && $0.01596$ && $0.31176 $&& $0.201226 $                                       \\
&&& $0.6$ && $0.01596$&& $0.305673 $&& $0.205161$ \\
&&& $0.9$ && $0.01596$ && $0.299981 $&& $0.208841 $ \\
&&& $1.2$ && $0.01596$ && $0.294644 $ && $0.212291$ \\
&&& $1.5$ && $0.01596$ && $0.289632 $ && $0.215532$ \\
\botrule
\end{tabular}
\end{table*}

\begin{table*}[t]
\centering
\caption{The numerical values of central density $\rho_c$, surface density $\rho_s$, central pressure $p_c$, mass, compactness and surface redshift for different values of $\beta$ for the compact star PSR J1614-2230.}
\label{table2}
\begin{tabular}{@{}cccccccccccccccc@{}}
\toprule
 $\beta$ && $\rho_c$&& $\rho_s$  &&$p_c$ & Maximum mass& $\mathcal{U}$ && $z_s(R)$ \\
&& $gm.cm^{-3}$ && $gm.cm^{-3}$ && $dyne.cm^{-2}$& km.\\
\hline
$0$ &&$2.57025 \times 10^{15}$ && $6.14818 \times 10^{14}$ && $7.97014 \times 10^{35}$& $2.90575 $ & $0.299871$ && $0.580629$\\
 $0.3$ && $2.49121 \times 10^{15}$ &&$5.93729 \times 10^{14}$ && $7.66728 \times 10^{35}$ &$2.81126$ & $0.29012$&& $0.543473$                                   \\
 $0.6$ && $2.41677 \times 10^{15}$&& $5.74176 \times 10^{14}$ && $7.39331 \times 10^{35}$&$2.72266 $ & $0.280976$&& $0.510913$\\
 $0.9$ && $2.34658 \times 10^{15}$ && $5.55986 \times 10^{14}$ && $7.14386 \times 10^{35}$& $2.63943$ & $0.272384$&& $0.48212$\\
 $1.2$ && $2.28031 \times 10^{15}$ && $5.39012 \times 10^{14}$ && $6.91543 \times 10^{35}$&$2.5611 $ & $0.264303$&& $0.456493$\\
 $1.5$ &&  $2.21764 \times 10^{15}$ && $5.23129 \times 10^{14}$ && $6.70514 \times 10^{35}$& $2.48727$ & $0.256684$&& $0.433507$\\
\botrule
\end{tabular}
\end{table*}

\begin{figure}[htbp]
    \centering
        \includegraphics[scale=.45]{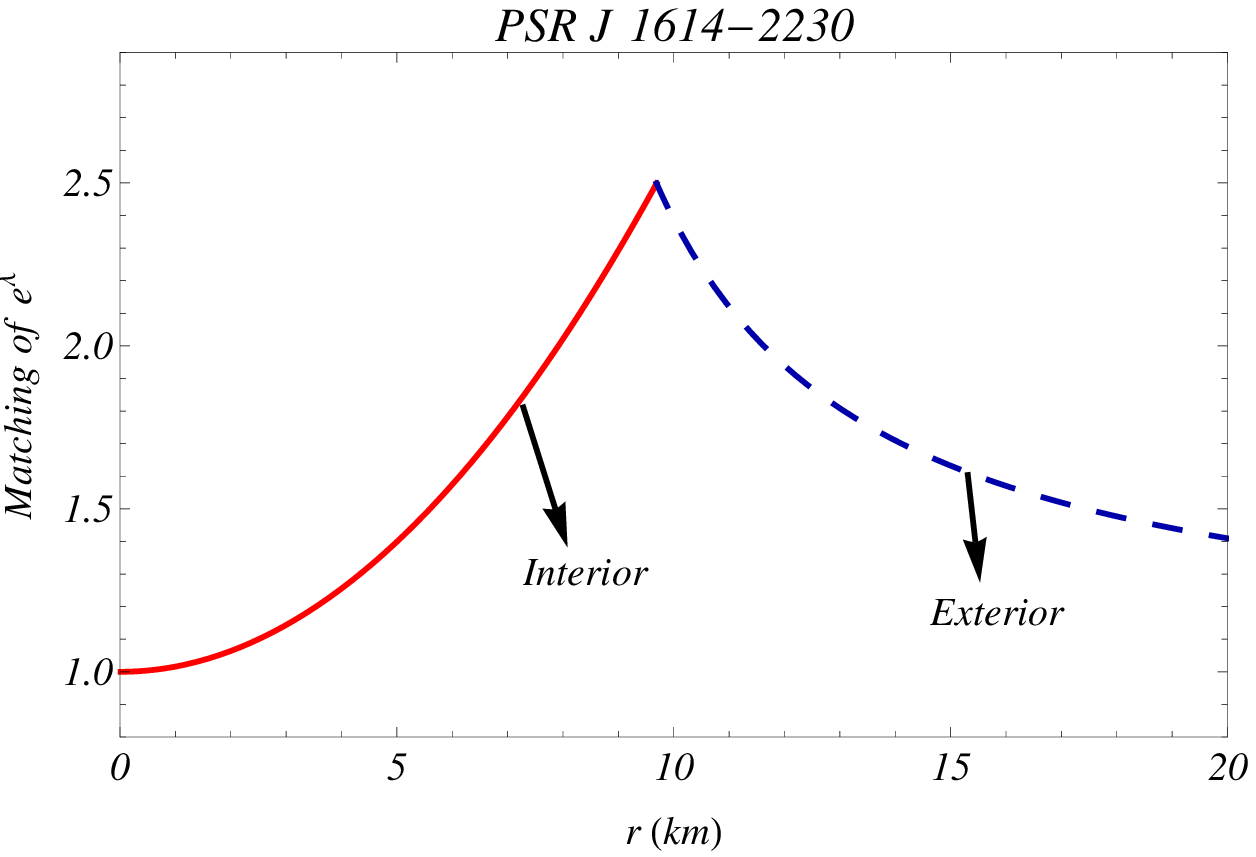}
        \includegraphics[scale=.45]{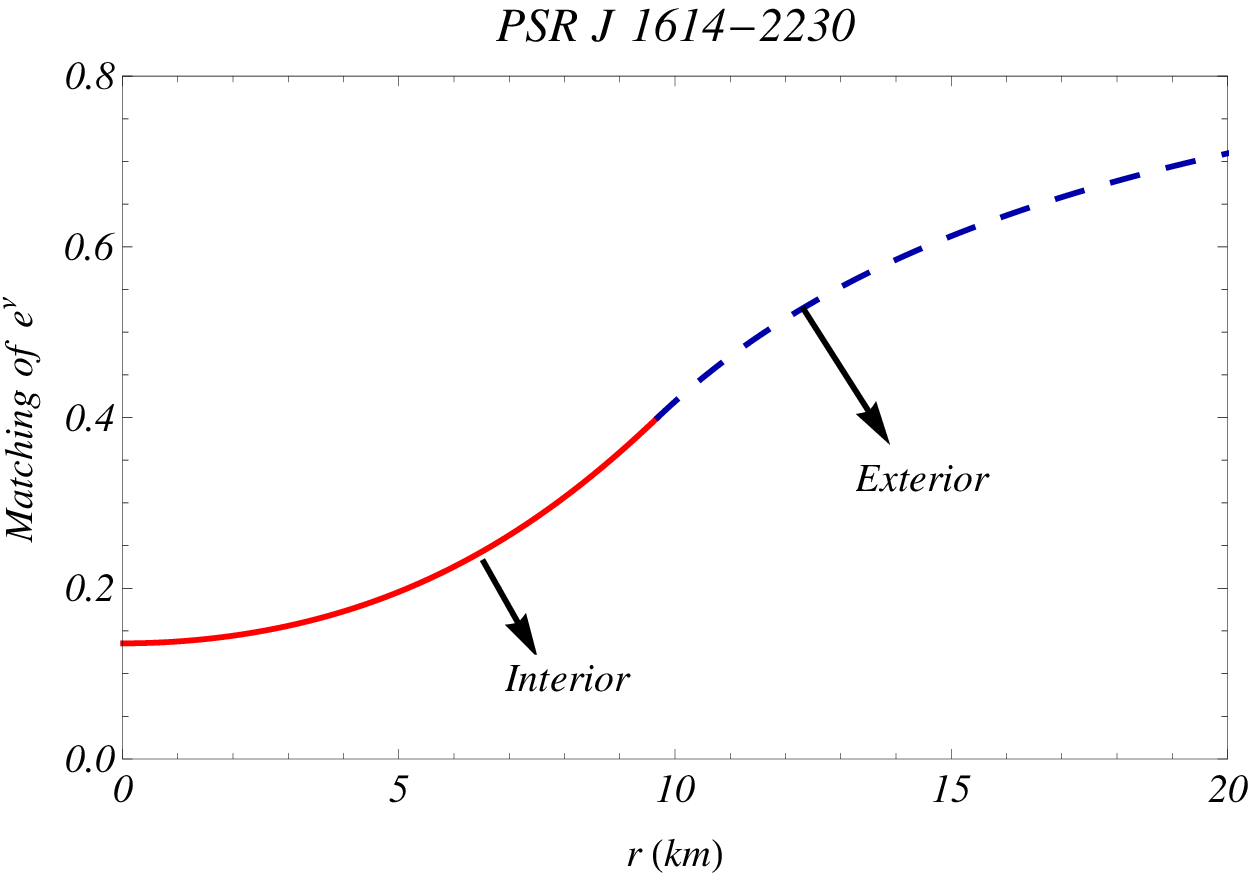}
       \caption{Both the metric coefficients are plotted against radius. The interior spacetime is smoothly matched with the exterior at the boundary.} \label{metric}
\end{figure}

At the center of the star, $e^{\lambda}=1$ and $e^{\nu}=((B-A)\cos 1+(A+B)\sin 1)^2$,
and the derivative of the metric coefficients are given by,
\begin{eqnarray*}(e^{\lambda})'&=& 2ar,\\(e^{\nu})'&=&2 a r \left(B \cos \sqrt{1 + a r^2} +
A \sin \sqrt{1 + a r^2}\right) \Big((B - A \sqrt{1 + a r^2}) \cos \sqrt{ 1 + a r^2} \nonumber\\&&+ (A + B \sqrt{1 + a r^2}) \sin \sqrt{1 + a r^2}\Big).\end{eqnarray*}
At the center of the star, the derivative of the metric coefficients vanish, so it can be concluded that the metric coefficients are regular at the center of the star. The nature of the metric coefficients are shown in Fig.~\ref{metric}. These plots also show the continuity of metric coefficients since the interior and exterior metric coefficients match at the stellar boundary.

\section{Analysis of physical characteristics}\label{pa}

We want to check the different physical properties of the model one-by-one in this section analytically
as well as with the help of graphical representation. In graphs, we vary the model parameter $\beta$
to check how it affects different quantities. Onwards in the manuscript, the color scheme is adopted as:
Red$\rightarrow$$\beta=0$ where $\beta=0$ corresponds to GR, Blue$\rightarrow$$\beta=0.3$, Green$\rightarrow$$\beta=0.6$,
Cyan$\rightarrow$$\beta=0.9$, Orange$\rightarrow$$\beta=1.2$ and Black$\rightarrow$$\beta=1.5$.

\subsection{Pressure and density profiles}

 The expressions of pressure and density are given in eqns.(\ref{p1})-(\ref{p2}) and their
 gradient can be found by simply taking the differentiation of these expressions of $\rho$ and $p$ as,
\begin{eqnarray}
\rho'&=& \frac{a^2 r}{8 (\beta + 2 \pi) (\beta + 4 \pi) Z^6}\bigg[-4 \Big\{(3 \beta + 8 \pi) (3 + a r^2) + \frac{
    \beta Z^2 \Psi_1}{\Psi_2}\Big\}+ \frac{Z}{\Psi_2^2}\Big\{4 (A^2 + B^2)Z f_1 \nonumber\\&&+
      2 (-A B (5 \beta + 16 \pi) Z^2 +
         B^2 Z f_2-f_2
         A^2 Z) \cos 2Z + \Big(-A^2 (5 \beta + 16 \pi) Z^2 \nonumber\\&&+
         B^2 (5 \beta + 16 \pi) Z^2 +
         4 A B Z f_2\Big) \sin 2Z\Big\}\bigg],\\
         p'&=&\frac{a^2 r}{8 (\beta + 2 \pi) (\beta + 4 \pi) Z^6}\bigg[-4 \Big\{\beta (3 + a r^2) + \frac{(3 \beta + 8 \pi)Z^2 \Psi_1}{\Psi_2}\Big\} - \frac{Z}{\Psi_2^2}\Big\{4 (A^2 + B^2) Z f_3 \nonumber\\&&+
      2 (-A B (\beta + 8 \pi) Z^2 +
         A^2 Z f_4 -
         B^2 Z f_4) \cos 2Z - \big(A^2 (\beta + 8 \pi) Z^2 -
         B^2 (\beta + 8 \pi) Z^2\nonumber\\&& +
         4 A B Z f_4\big) \sin 2Z\Big\}\bigg],
\end{eqnarray}
where $f_i$'s are functions of `r' given by,
\begin{eqnarray*}
f_1&=&\beta + 2 \pi (2 + a r^2),\\
f_2&=&\beta - a (\beta + 4 \pi) r^2,\\
f_3&=&\beta + 4 \pi + 2 a (\beta + 3 \pi) r^2,\\
f_4&=&3 \beta + 8\pi + a (\beta + 4 \pi) r^2.
\end{eqnarray*}

\begin{figure}[htbp]
    \centering
        \includegraphics[scale=.45]{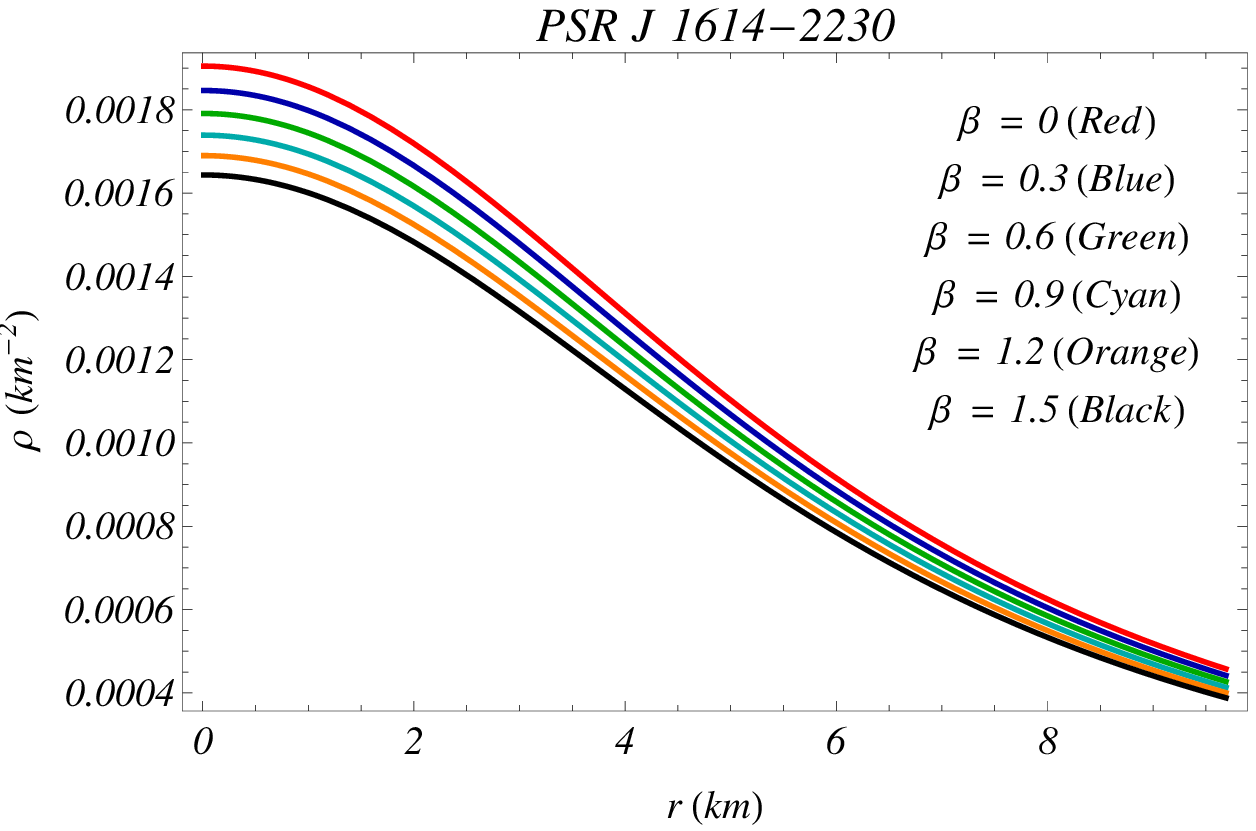}
        \includegraphics[scale=.45]{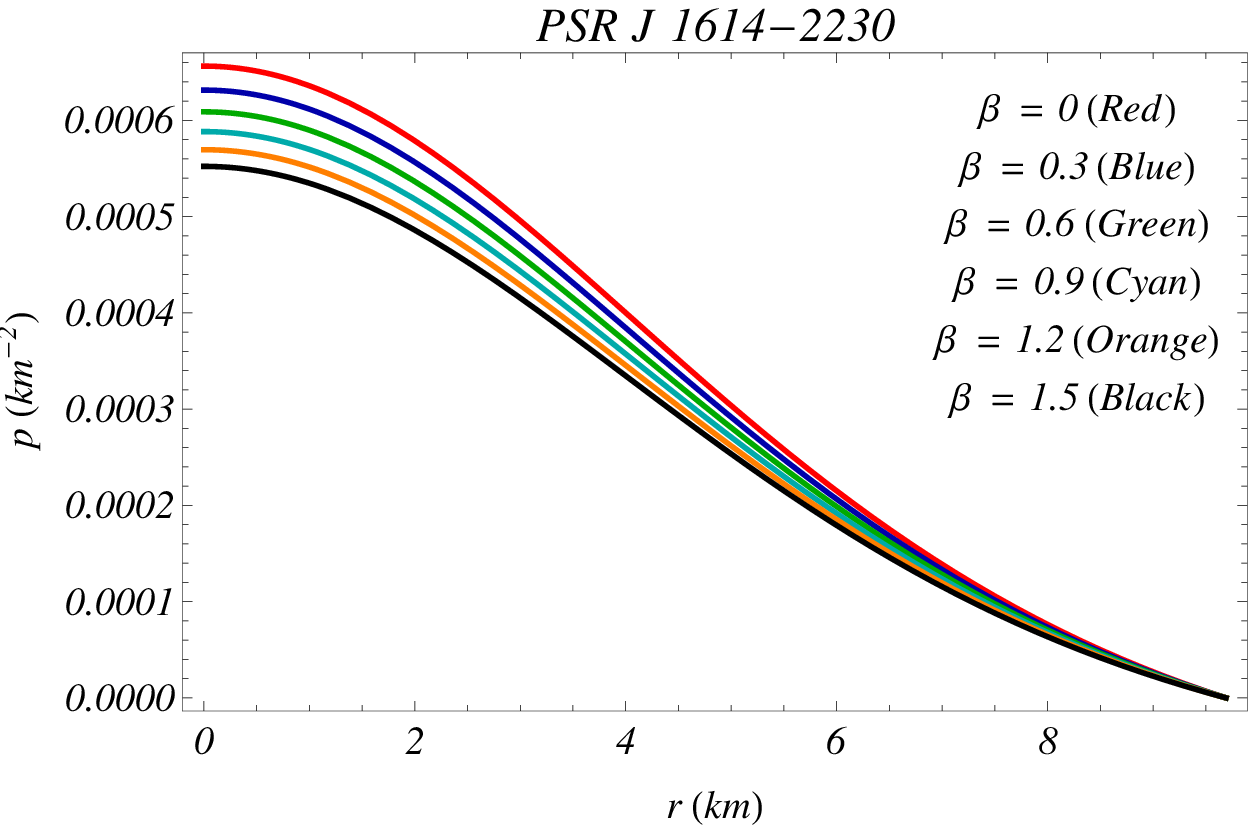}
       \caption{(Left) matter density and (right) pressure are plotted against radius for different values of the coupling constants mentioned in the figure.}\label{rho1}
\end{figure}
\begin{figure}[htbp]
    \centering
        \includegraphics[scale=.45]{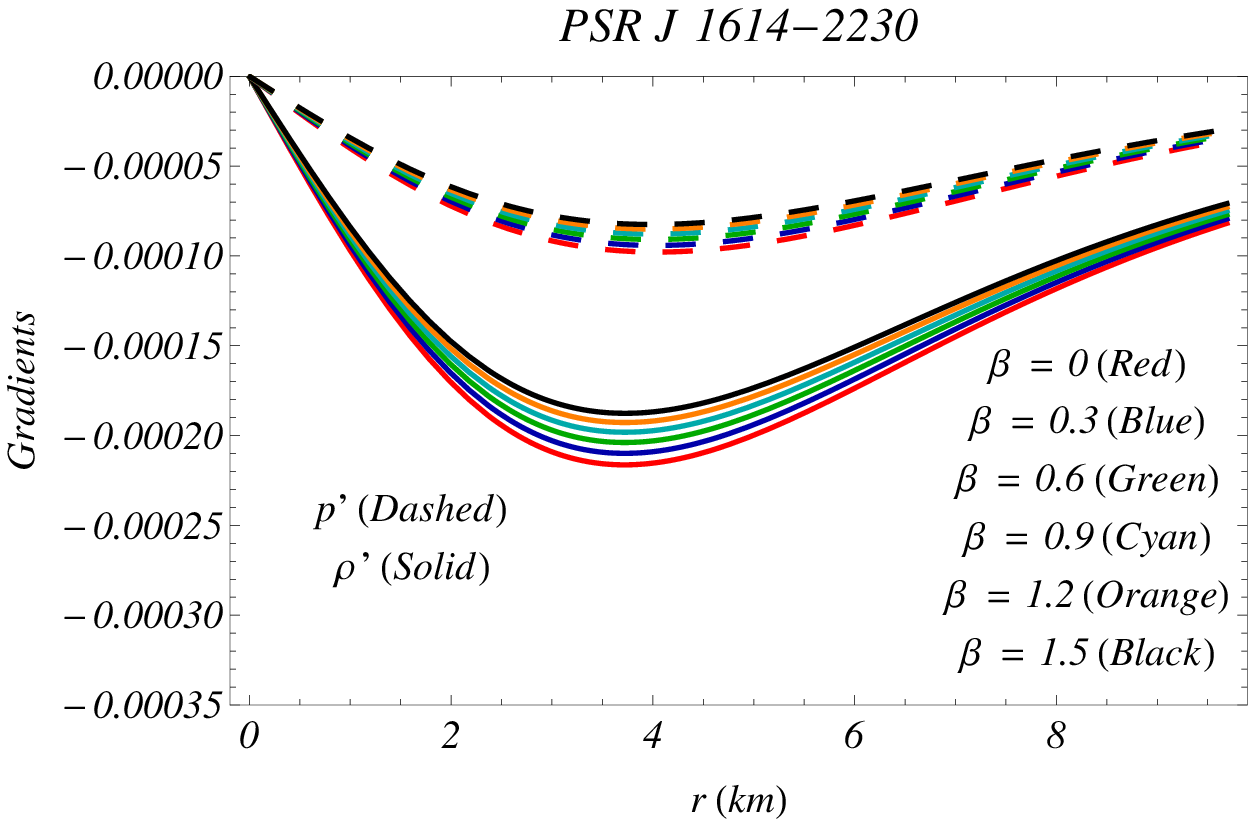}
        \includegraphics[scale=.45]{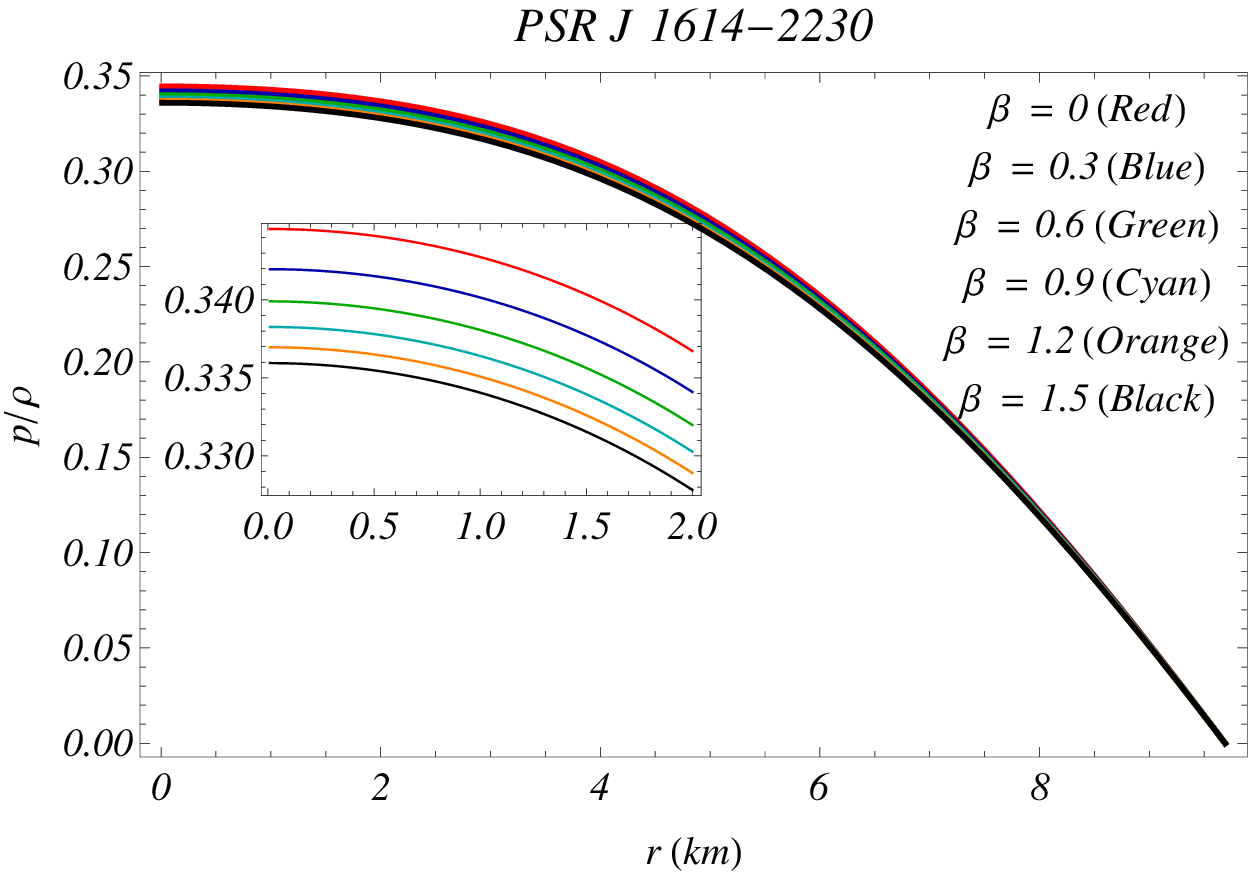}
       \caption{(Left) the pressure and density gradients are shown against radius and (right) the ratio of pressure to the density is shown against radius.\label{deri}}
\end{figure}
\begin{figure}[htbp]
    \centering
        \includegraphics[scale=.45]{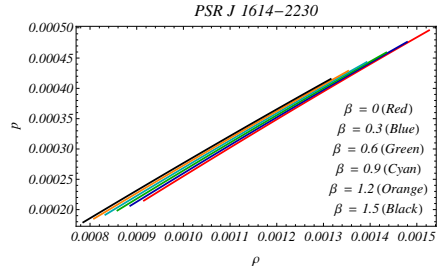}
       \caption{The pressure versus density.}\label{eos}
\end{figure}

The obtained expressions of density and pressure of the considered
model are plotted in Fig.~\ref{rho1}. It is observed from these
plots that $\rho$ and $p$ admit the general behavior for stellar
configurations, i.e., are maximum at the center and decrease
monotonically towards the boundary. These figures also depict that
the density and pressure of stellar fluid constrained with the Finch-Skea model decrease with an increase in the value of $\beta$.
Moreover, their derivative functions are negative within the stellar
interior as shown in left panel of Fig.~\ref{deri}. These plots
ensure that the density and pressure are decreasing functions with
respect to radial coordinate within the domain of interior
configuration. The ratio of pressure to density is known as equation
of state parameter. For the current matter, this parameter is
positive and attain a maximum value of $0.35$ corresponding to
$\beta=0$ while decreases with an increase in $\beta$. Since $p/\rho$ lies in the range (0,\,1) for different values of the coupling constant $\beta$, it indicates that the underlying matter distribution is non-exotic in nature. In Fig.~\ref{eos}, the pressure is plotted against
density and the figure indicates that they obey a linear relationship.

\subsection{Energy conditions}

The fulfilment of energy conditions ensure the compatibility of geometry and matter. We check these conditions for the
assumed configuration to verify the compatibility of Finch-Skea geometry with perfect fluid matter as well as the acceptability of obtained solutions.
These conditions are defined below
\begin{itemize}
\item Null energy condition: $\rho+ p\geq 0$,
\item Weak energy condition: $\rho\geq 0$, $\rho+ p\geq 0$,
\item Strong energy condition: $\rho+ 3p\geq 0$,
\item Dominant energy condition: $\rho\geq|p|$,
\end{itemize}

\begin{figure}[htbp]
    \centering
        \includegraphics[scale=.45]{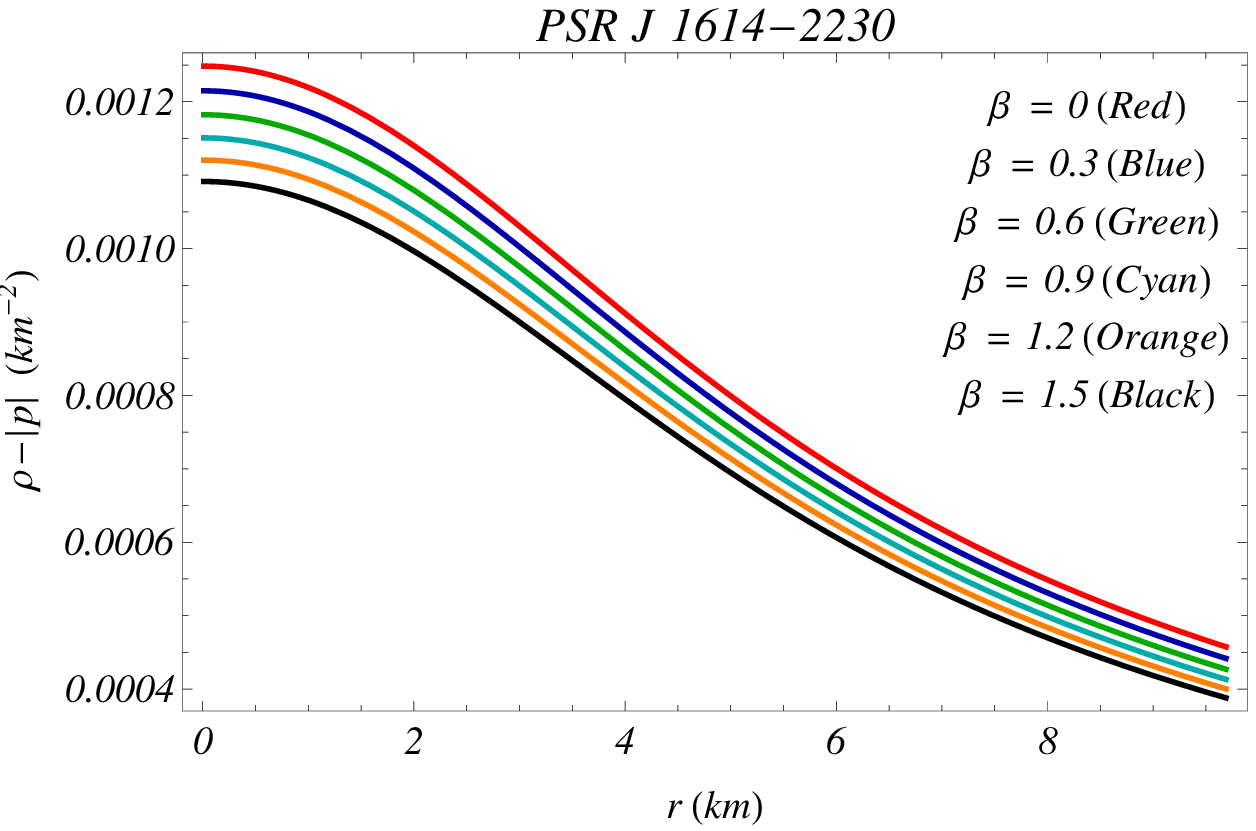}
        \includegraphics[scale=.45]{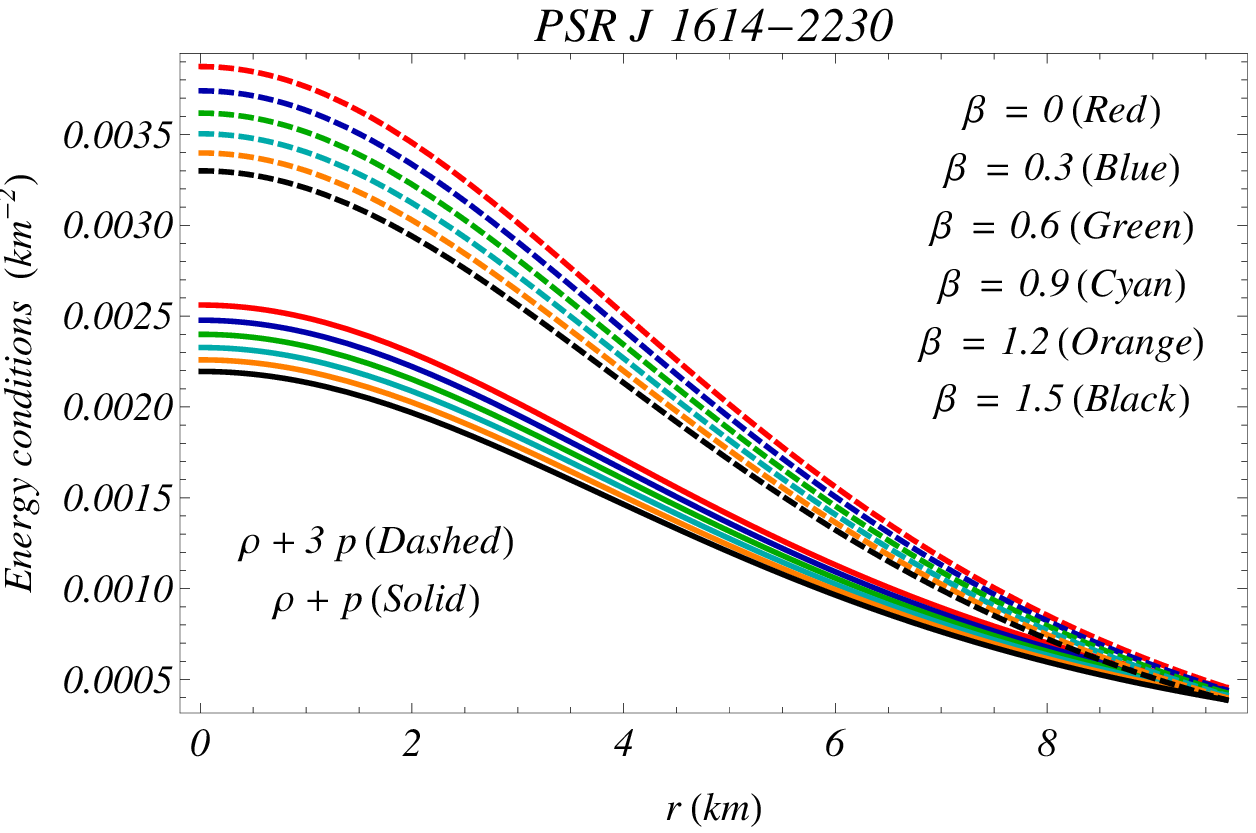}
       \caption{The energy conditions are plotted against radius inside the stellar interior \label{ec1}}
\end{figure}
From the plots given in Fig.~\ref{ec1}, it is obvious that the
null, weak, strong and dominant energy conditions are satisfied implying that the solutions are physically acceptable.

\subsection{Mass and associated factors}

\begin{figure}[htbp]
    \centering
        \includegraphics[scale=.45]{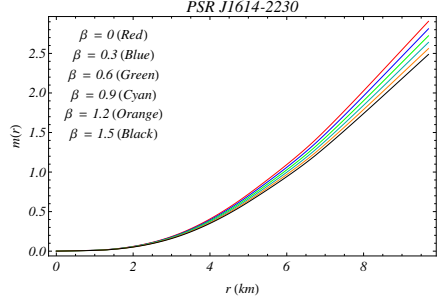}
       \caption{The mass function is plotted against radius inside the stellar interior}\label{mass1}
\end{figure}

The mass $m(r)$, compactness $\mathcal{U}$ and surface redshift parameter $z_{s}(R)$ for stellar structure are defined as follows:
\begin{eqnarray}\nonumber
m(r)&=&4\pi\int_{0}^{r}\rho r^{2}=\frac{8\pi}{8\pi+3\beta}m^{\text{eff}}+\frac{4\pi \beta}{8\pi+3\beta}\int_0^r pr^2dr,\\\nonumber
\mathcal{U}&=&\frac{m(R)}{R},\\\nonumber
z_{s}(R)&=&\left(1-2\mathcal{U}\right)^{-\frac{1}{2}}-1.
\end{eqnarray}
Where $m^{\text{eff}}$ is the effective gravitational mass defined by $m^{\text{eff}}=\int_0^r 4\pi \rho^{\text{eff}}r^2dr=\frac{ar^3}{2(1+ar^2)}$.
For stellar structures, the value of mass function is zero at the center of the star and maximum at the boundary of star. From literature it is known that the upper bound of mass for a white
dwarf is $1.4M_{\odot}$ and for a neutron star, the upper bound for mass is $3M_{\odot}$. The profile of mass function for different values of $\beta$ is shown in Fig.~\ref{mass1}. From the Fig.~\ref{mass1}, it is observed that the mass function takes lower value with the increasing value of $\beta$. The mass function is regular and monotonic increasing function of $r$. In General Relativity, compactness factor has an upper bound of $\frac{4}{9}$ \cite{Buchdahl:1959zz} while surface redshift $z_s(R)$ has a maximum limit $z_s(R)\leq2$
for isotropic distributions \cite{Buchdahl:1959zz,Straumann:1984xf,Boehmer:2006ye}. The numerical values of mass, compactness factor $\mathcal{U}$ and surface redshift $z_s(R)$ have been shown in table~\ref{table2}. The table~\ref{table2} indicates
that compactness factor and surface redshift lie within the expected range.

\section{Stability analysis}\label{stability}

Till now, we have presented the physical properties of a new
compact star model having the geometry described by Finch-Skea
metric potential with perfect fluid matter as well as the unknown
constants are obtained by using mass and radius of PSR~J 1614-2230.
The analysis will be incomplete if we do not discuss its stability
as well. Stability of any stellar structure guarantees that the
equilibrium of the system will not be disturbed.

\subsection{Equilibrium of system}

\begin{figure}[htbp]
    \centering
        \includegraphics[scale=.45]{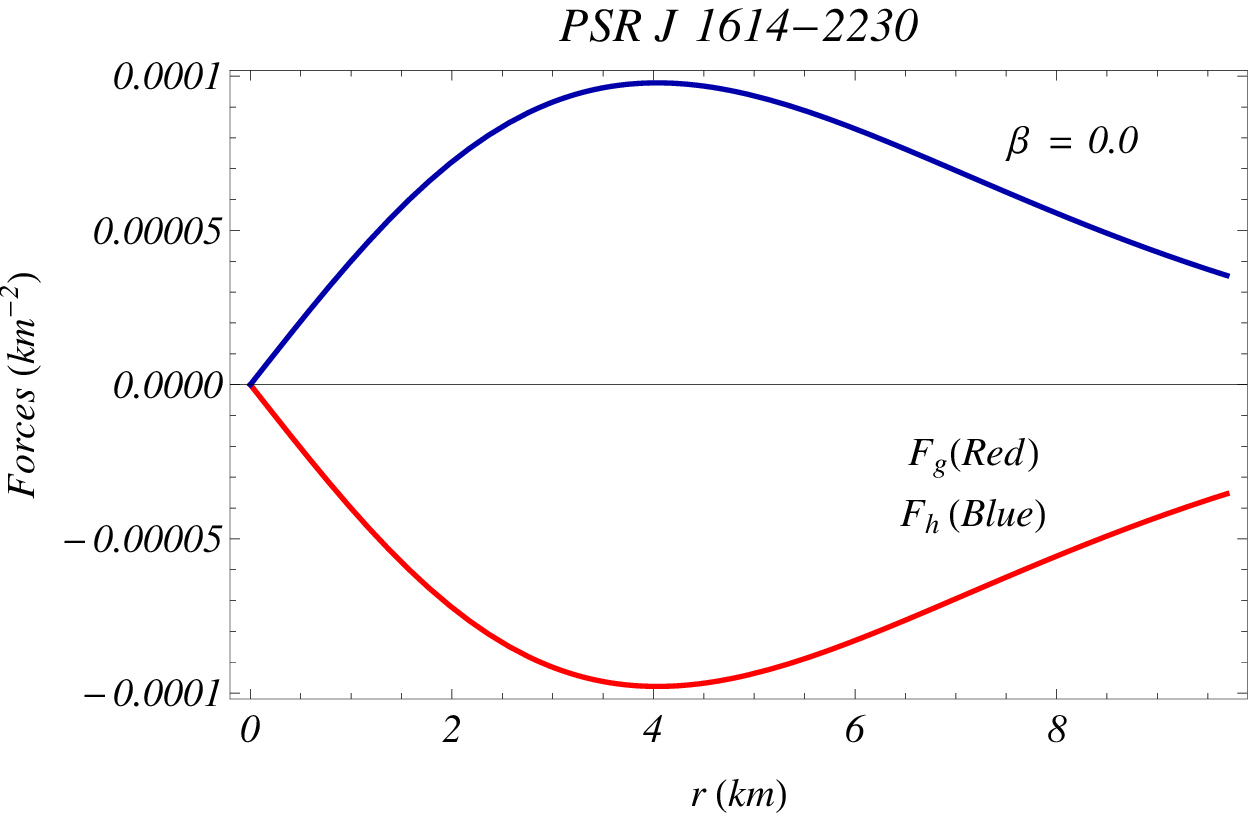}
        \includegraphics[scale=.45]{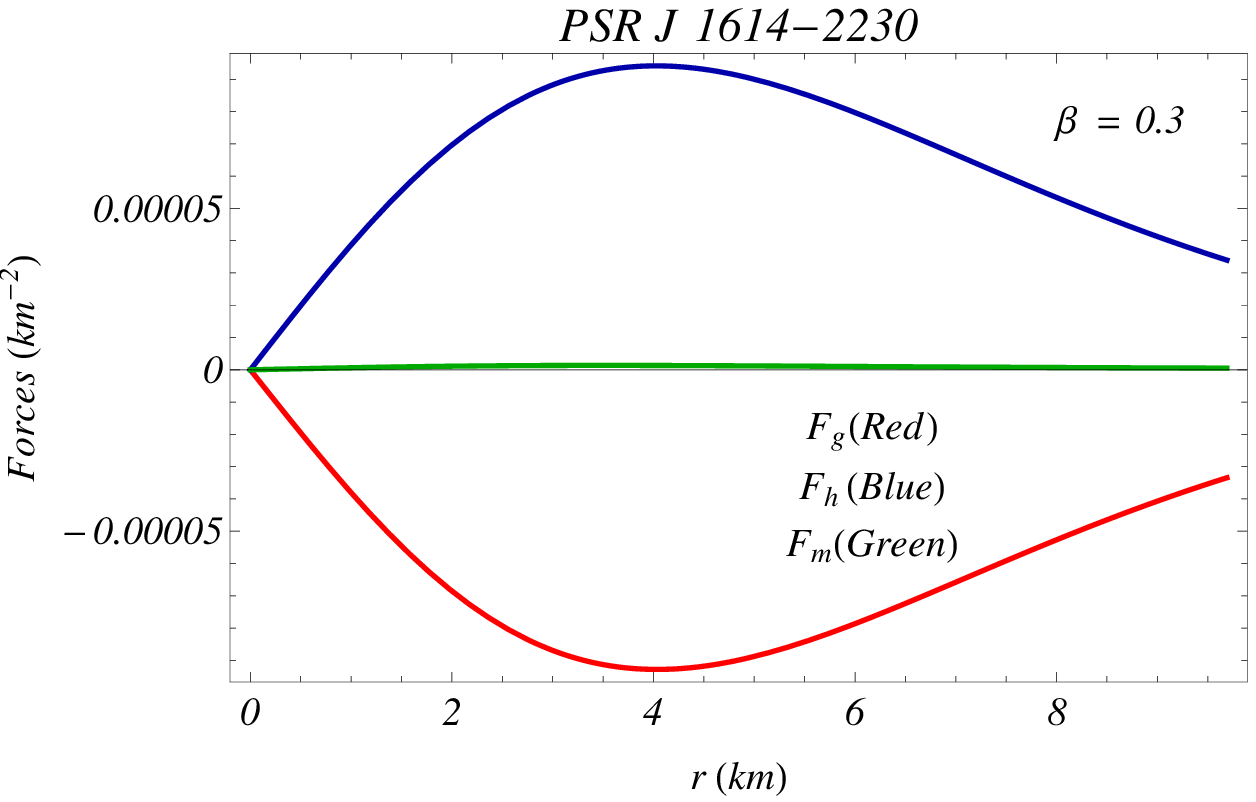}
        \includegraphics[scale=.45]{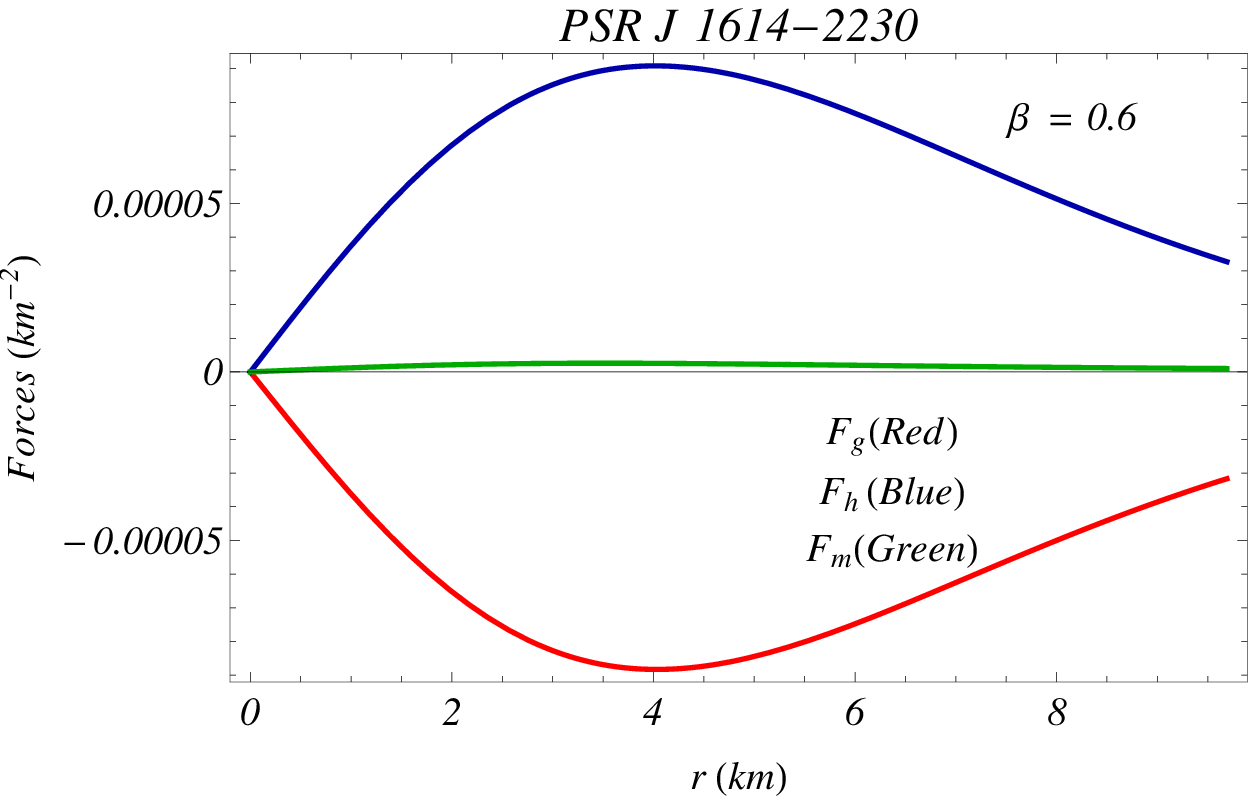}
        \includegraphics[scale=.45]{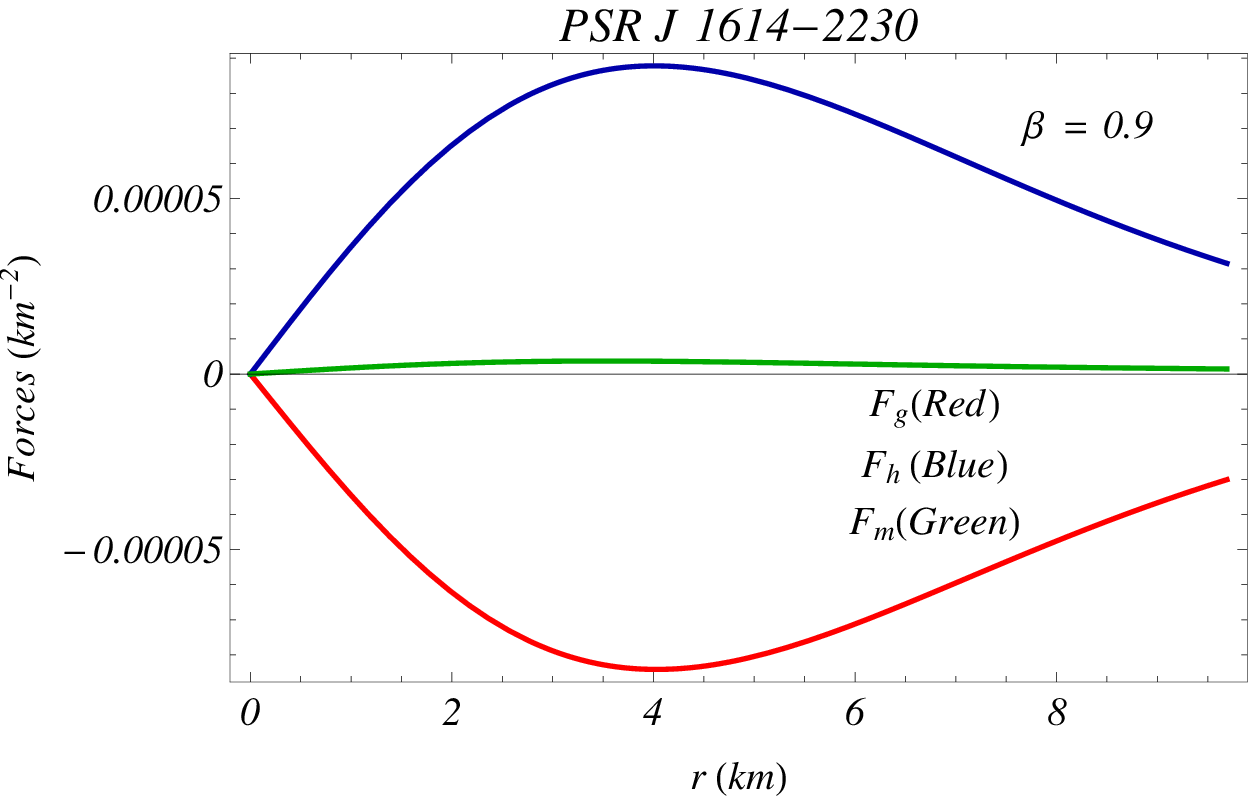}
        \includegraphics[scale=.45]{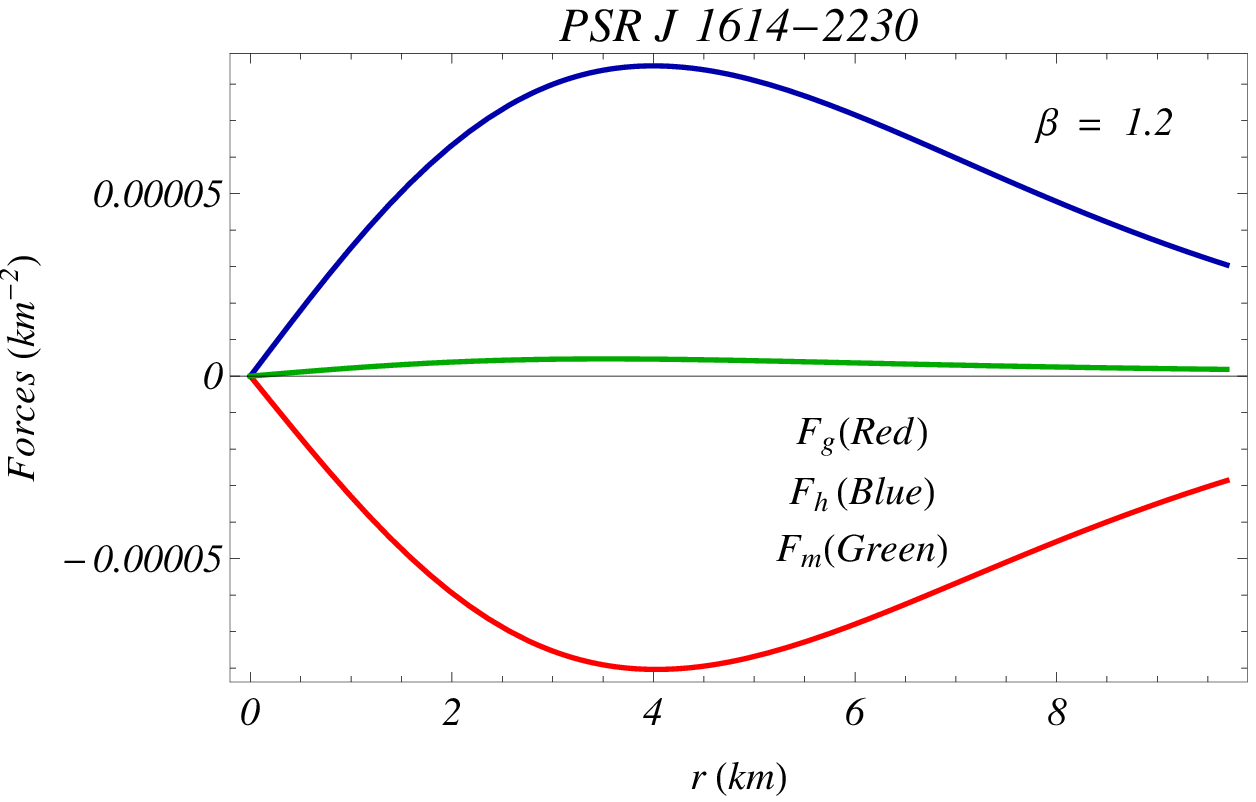}
        \includegraphics[scale=.45]{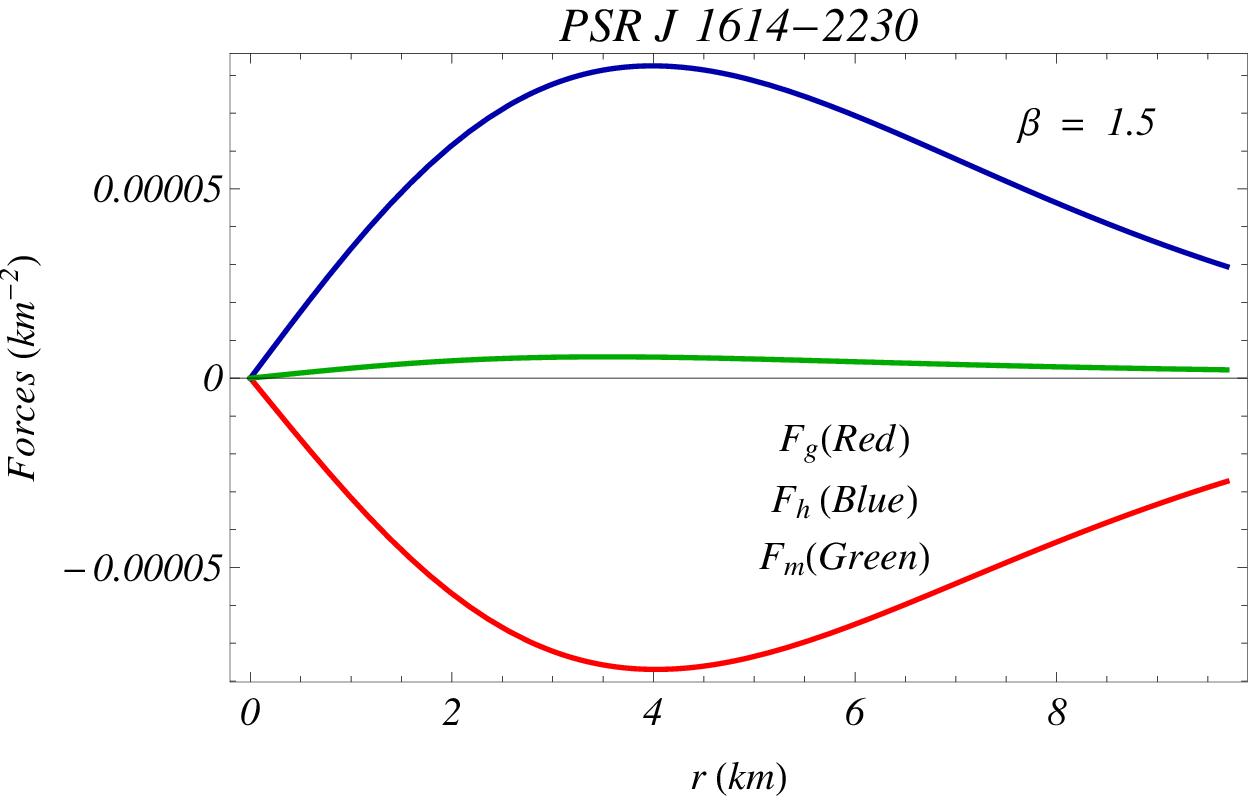}
       \caption{Different forces acting on the system are plotted against radius inside the stellar interior for different values of $\beta$.\label{tov1}}
\end{figure}

First we want to check  whether our proposed system is in equilibrium under different forces or not. This is done by formulating the TOV equation given
below:
\begin{eqnarray*}
F_{g}&+&F_{h}+F_{m}=0,
\end{eqnarray*}
where the expressions for $F_g$, $F_h$ and $F_m$ are given by,
\begin{eqnarray*}
F_{g}&=&\frac{-1}{2(\rho+p)}\frac{d}{dr}\left(\ln(B-AZ)\cos
Z+(A+BZ)\sin Z\right)^{2},\\\nonumber
F_{h}&=&-\frac{dp}{dr},\\\nonumber
F_{m}&=&\frac{\beta}{8\pi+2\beta}\left(\frac{dp}{dr}-\frac{d\rho}{dr}\right).
\end{eqnarray*}
Here $F_{g}$, $F_{h}$ and $F_{m}$ correspond to gravitational force,
hydrostatic force and force due to modified gravity, respectively.
All plots in Fig.~\ref{tov1} indicate that the stellar system is
in equilibrium phase in the considered scenario for different values
of $\beta$.

\subsection{Causality condition}

\begin{figure}[htbp]
    \centering
        \includegraphics[scale=.45]{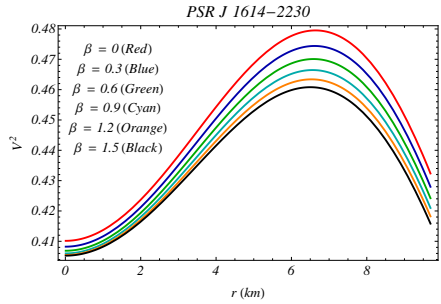}
       \caption{ The square of the sound velocity is plotted against radius inside the stellar interior.}\label{sv1}
\end{figure}

Causality and stability of a system are related in a sense that the system
will not be stable if and only if there will be a mode in the system which is moving faster than the speed of
light. Hence it is important to check the causality condition for a stellar structure to ensure its stability.
For this purpose, we consider the speed of sound as $V^{2}=\frac{dp}{d\rho}$. We want to verify whether it is between
zero and one or not as the causality condition for a model will be satisfied if $0<\frac{dp}{d\rho}<1$. The plot in Fig.~\ref{sv1} ensures that our proposed model of compact star obeys causality and hence it is stable.

\subsection{Adiabatic index}
\begin{figure}[htbp]
    \centering
        \includegraphics[scale=.45]{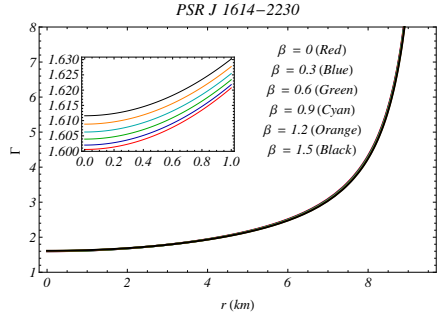}
       \caption{The relativistic adiabatic index is plotted against radius inside the stellar interior. \label{gamma1}}
\end{figure}
Adiabatic index is the main ingredient in dynamical stability because the stability of compact stellar structure
is mainly depending upon the interior fluid's EoS as well as role of relativistic field and this
quantity serves as a bridge between the relativistic structure and EoS of internal matter in a stellar structure.
It indicates the stiffness of the equation of states at a given density and is defined by
\begin{eqnarray}\nonumber
\Gamma=\frac{p+\rho}{p}\frac{\partial p}{\partial \rho}.
\end{eqnarray}
For stable configurations the value of $\Gamma$ should exceed $\frac{4}{3}$ and it can be observed from
Fig.~\ref{gamma1} that our proposed model is stable for the assumed geometrical constraints and matter because $\Gamma>\frac{4}{3}$ everywhere within the stellar configuration for different values of $\beta$.

\section{Discussions and Concluding Remarks}\label{conclusion}

The human curiosity to explore the space mysteries lead to many
observations. The researchers also proposed many theoretical ideas
to unveil the unknown issues like dark energy, dark
matter and other various astrophysical and cosmic issues. In
this article, we studied the physical features of a compact star
model and its stability in the framework of $f(R,T)$ theory. Mainly,
we impose two constraints to the spherically symmetric perfect fluid
distribution to solve the corresponding field equations, firstly,
Finch-Skea {\em ansatz} and secondly compact star candidate PSR~J
1614-2230 to find the unknown constants. After that, we investigate
various characteristics of proposed compact star model as well as
their behavior corresponding to change in coupling parameter $\beta$
via graphs.

The density, pressure and mass profiles show the usual behavior of
these functions for stellar structures. However, density and
pressure tend to decrease with an increase in $\beta$. All the energy conditions
are satisfied which implies the comparability of matter and geometry. The
gravitational, hydrostatic and coupling forces due to modified
gravity balance themselves to keep the system in equilibrium.
Furthermore, the causality condition is satisfied inside the boundary and relativistic adiabatic index for
stable configurations is hold.

We have shown the regularity of metric functions at the center both analytically and graphically. The pressure to density ratio
is between $0$ and $1$ implying realistic matter distribution. It is
important to mention that we also plot the graphs for $\beta=0$
which gives a comparison of the present work in $R+2\beta T$ scenario
with General Relativity. From literature \cite{Moraes:2015uxq,Sharif:2018khl}, we see that the increase in
coupling parameter for this model tend to enhance the mass function
while in present framework we see the reverse nature of the mass function, i.e., with the increasing value of $\beta$, the numerical value of mass decreases. In view of all above discussion we conclude
that the present scenario provides a suitable theoretical framework
to model the quark stars in the context of modified gravity.

\section*{Acknowledgments}
 PB is thankful to IUCAA, Government of India for providing visiting associateship.

\bibliographystyle{unsrt}
\bibliography{fs}

\end{document}